\documentclass[conference]{IEEEtran}
\IEEEoverridecommandlockouts

\usepackage{cite}
\usepackage{amsmath,amssymb,amsfonts}
\usepackage{algorithmic}
\usepackage{graphicx}
\usepackage{textcomp}
\usepackage{xcolor}
\usepackage{url}
\usepackage{booktabs}
\usepackage{listings}
\def\BibTeX{{\rm B\kern-.05em{\sc i\kern-.025em b}\kern-.08em
    T\kern-.1667em\lower.7ex\hbox{E}\kern-.125emX}}
\begin{document}

\title{ArchBench: Benchmarking Generative-AI \\for Software Architecture Tasks
}

\author{
\IEEEauthorblockN{Bassam Adnan, Aviral Gupta, Sreemaee Akshathala, Karthik Vaidhyanathan}
\IEEEauthorblockA{
\textit{Software Engineering Research Center (SERC)} \\
\textit{International Institute of Information Technology Hyderabad (IIIT-H)} \\
Hyderabad, India \\
\{bassam.adnan, aviral.gupta, sreemaee.akshathala\}@research.iiit.ac.in, karthik.vaidhyanathan@iiit.ac.in
}
}

\maketitle

\begin{abstract}

Benchmarks for large language models (LLMs) have progressed from snippet-level function generation to repository-level issue resolution, yet they overwhelmingly target implementation correctness. Software architecture tasks remain under-specified and difficult to compare across models, despite their central role in maintaining and evolving complex systems. We present ArchBench, the first unified platform for benchmarking LLM capabilities on software architecture tasks. ArchBench provides a command-line tool with a standardized pipeline for dataset download, inference with trajectory logging, and automated evaluation, alongside a public web interface with an interactive leaderboard. The platform is built around a plugin architecture where each task is a self-contained module, making it straightforward for the community to contribute new architectural tasks and evaluation results. We use the term LLMs broadly to encompass generative AI (GenAI) solutions for software engineering, including both standalone models and LLM-based coding agents equipped with tools. Both the CLI tool and the web platform are openly available to support reproducible research and community-driven growth of architectural benchmarking.

\end{abstract}


\section{Introduction}
\label{sec:introduction}
\begin{figure*}[t]
    \centering
    \includegraphics[width=\textwidth]{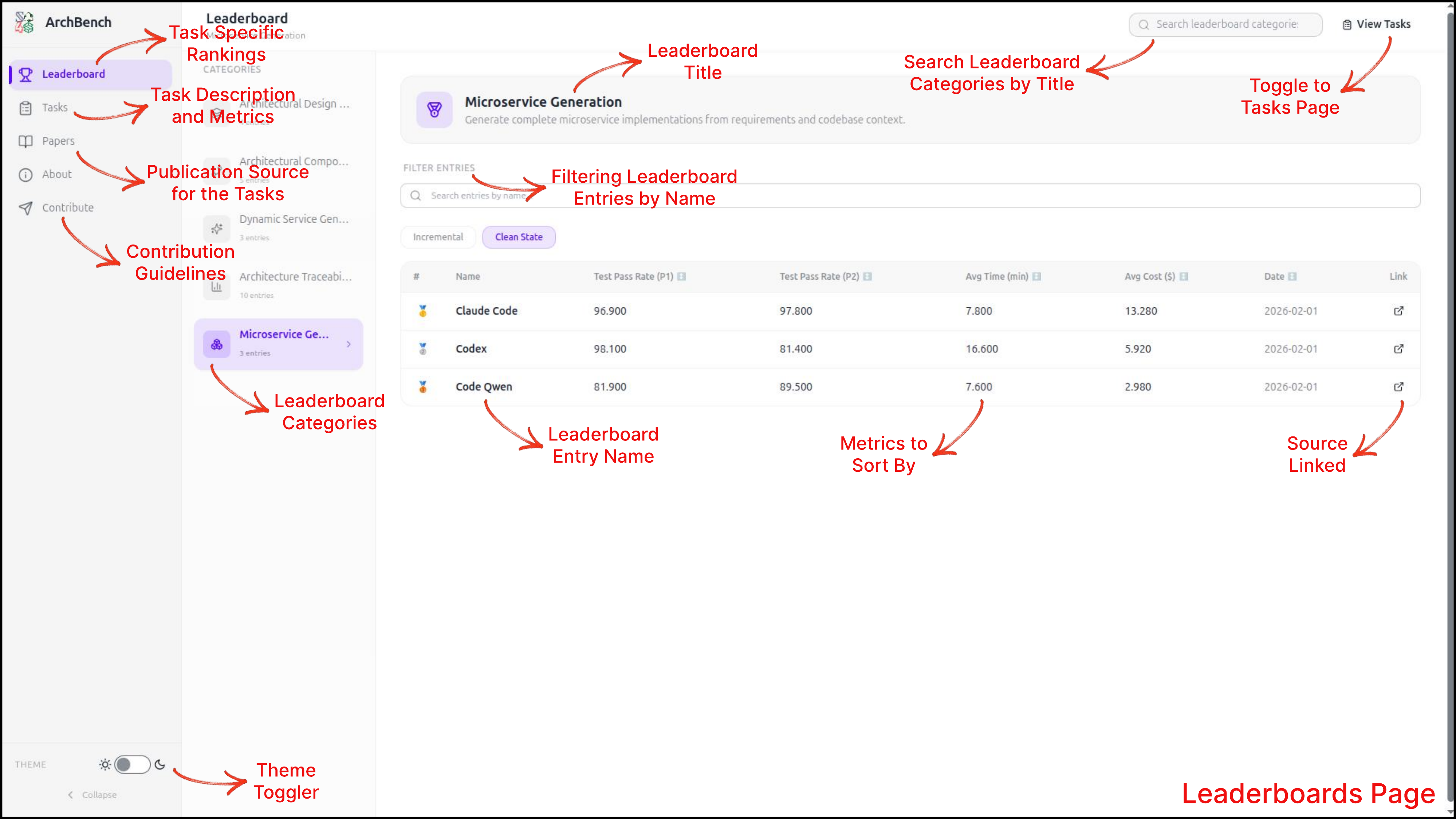}
    \caption{Annotated screenshot of the ArchBench web interface. Circled elements highlight the platform's key sections: the leaderboard for comparing model performance across tasks, task descriptions with evaluation metrics, source papers for each dataset, and contribution guidelines for community submissions.}
    \label{fig:archbench-ui}
\end{figure*}

Software architecture is foundational to software quality. Decisions about modularization, dependency management, and system evolution determine long-term maintainability, and unlike implementation bugs, architectural flaws compound quietly until they dominate engineering effort \cite{ Li2015TD, Siebra2012TD, Verdecchia2018ATD, Rios2020TD, Verdecchia2020ATD, Fuchs2025ICSA}. As LLMs are increasingly deployed as architectural advisors and refactoring tools, evaluating their architectural reasoning has become critical. Yet current benchmarks focus overwhelmingly on implementation correctness, ranging from snippet-level function generation to repository-level issue resolution, with no standardized evaluation targeting software architecture tasks. A recent multivocal literature review on GenAI for software architecture~\cite{ESPOSITO2026} confirms this gap, finding that rigorous testing of GenAI outputs is typically missing, architecture-specific datasets are scarce, and sound evaluation frameworks are absent. This means architectural capabilities cannot be compared, tracked, or reliably improved across models. While the broader AI and software engineering communities have benefited greatly from shared datasets and common evaluation platforms that accelerate progress, the software architecture community currently lacks such infrastructure.

We address this gap with ArchBench, the first unified platform for benchmarking LLM capabilities on software architecture tasks. ArchBench provides an extensible framework for aggregating architectural tasks from published research, each with datasets, task-specific metrics, and ground-truth references. The platform consists of a command-line tool implementing a standardized evaluation pipeline and a public web interface with a leaderboard. New tasks can be contributed as self-contained plugins without modifying the core framework, and new evaluation results are submitted via pull requests.
Our work makes two primary contributions:
\begin{itemize}
  \item \textbf{A centralized benchmarking platform for software architecture tasks}: We present the first unified platform for aggregating datasets from published research on GenAI for software architecture. The platform currently supports five tasks drawn from recent studies that evaluate LLM capabilities on architectural problems, and is designed for extensibility, allowing the community to contribute additional tasks over time.
  \item \textbf{A standardized evaluation pipeline}: We provide a command-line tool that handles dataset download, LLM inference with full trajectory logging, and automated metric computation, producing standardized output formats that enable consistent, reproducible comparison across models.
\end{itemize}

The remainder of this paper is structured as follows. Section~\ref{sec:technical} details the technical architecture of the platform and its current tasks. Section~\ref{sec:usage} describes usage workflows with a walkthrough. Section~\ref{sec:discussion} discusses limitations and future directions.
\section{ArchBench Overview}
\label{sec:technical}

\begin{figure}[t]
    \centering
    \includegraphics[width=0.9\linewidth]{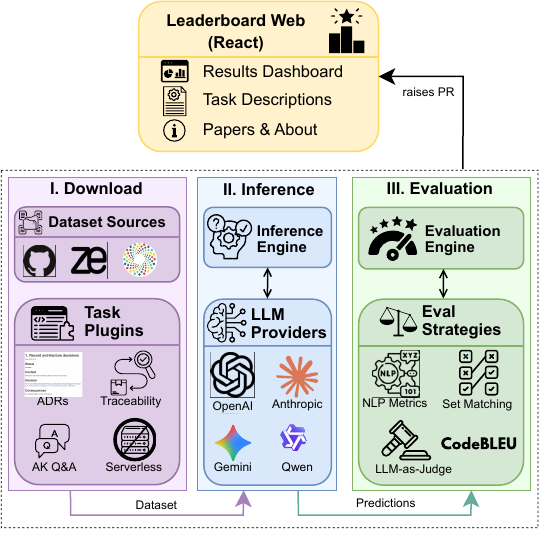}
    \caption{ArchBench platform architecture showing the three pipeline stages (Download, Inference, Evaluation) and the leaderboard web interface.}
    \label{fig:architecture}
\end{figure}

ArchBench consists of a Python command-line tool and a public web interface. The CLI implements a three-stage pipeline (download, inference, and evaluation) while the web interface provides a leaderboard for browsing results. Fig.~\ref{fig:architecture} shows the overall architecture. At the core is a plugin system where each task is a self-contained module providing dataset loading, prompt templates, response parsing, and metric definitions. New tasks can be added without modifying the framework.

\noindent \textbf{Stage I: Download.}
Datasets are automatically fetched from external sources (GitHub, Zenodo, HuggingFace) on first use and cached locally. Each task plugin defines its own data loader, handling format-specific parsing.

\noindent \textbf{Stage II: Inference.}
The inference engine assembles prompts from task-specific templates, dispatches them to LLM providers (OpenAI, Anthropic, and others via a uniform provider interface), and parses raw responses into structured predictions. Full interaction trajectories, including prompts, raw responses, token usage, and latency, are logged for reproducibility.

\noindent \textbf{Stage III: Evaluation.}
The evaluation engine validates predictions, computes task-appropriate metrics, and generates standardized reports. Metrics are selected per task: NLP similarity metrics (ROUGE, BLEU, BERTScore) for text generation tasks~\cite{Zhang2020BERTScore}, set-based precision/recall/F1 for structured outputs like traceability links, code-level metrics (CodeBLEU, test pass rates) for code generation, and optionally LLM-as-judge for qualitative assessment~\cite{Gu2025LLMJudge, Dhar2024ICSA}.

\noindent \textbf{Web Interface.}
A React application publicly available at \url{https://www.sabench.com/} displays results across all tasks. Fig.~\ref{fig:archbench-ui} shows an annotated screenshot of the interface. The sidebar lists leaderboard categories for each task, while the main panel shows a sortable table of model entries with task-specific metrics, links to source publications, and sub-view toggles for tasks with multiple evaluation settings.

\subsection*{Supported Tasks}
\label{sec:tasks}

ArchBench currently includes five tasks drawn from recent software architecture research. Each task was integrated with explicit permission from its original authors. Table~\ref{tab:tasks} provides a summary; we briefly describe each below. The selection of tasks is guided by the taxonomy of GenAI applications in software architecture identified by Esposito et al.~\cite{ESPOSITO2026}, which categorizes LLM use across areas such as architectural decision-making, requirements-to-architecture transformation, modularization, and architectural knowledge extraction. The current five tasks cover a subset of these categories, and ArchBench is designed to progressively incorporate tasks spanning the full taxonomy as suitable datasets and evaluation methodologies become available.

\begin{table}[t]
\caption{Tasks currently supported by ArchBench.}
\label{tab:tasks}
\centering
\small
\begin{tabular}{@{}p{2.2cm}p{2.8cm}p{2.5cm}@{}}
\toprule
\textbf{Task} & \textbf{Description} & \textbf{Metrics} \\
\midrule
ADR Generation & Generate architectural decision records from contexts & ROUGE, BLEU, METEOR, BERTScore \\
\addlinespace
Serverless Component Gen & Generate FaaS functions from specifications & Test pass rate, CodeBLEU, Complexity \\
\addlinespace
Dynamic Service Gen & Generate IoT services from task descriptions & CodeBERTScore (P/R/F1/F3) \\
\addlinespace
Traceability Link Recovery & Recover trace links between SAD and code & Precision, Recall, F1 \\
\addlinespace
Microservice Generation & Generate microservices from requirements & Test pass rate, SLOC, Complexity \\
\bottomrule
\end{tabular}
\end{table}

\noindent \textbf{ADR Generation.} Given an architectural context describing a design problem, the model must generate an Architecture Decision Record (ADR) capturing the decision and its rationale. The dataset contains 1000 architectural documents sourced from open-source projects \cite{Dhar2024ICSA}. Evaluation uses standard NLP similarity metrics (ROUGE, BLEU, METEOR) alongside BERTScore for semantic similarity.

\noindent \textbf{Architectural Component Generation.} Given a specification and surrounding codebase context, the model must generate complete serverless (FaaS) functions. The Wonderless dataset provides 10 functions across 4 repositories\cite{Arun2025ICSA, Eismann2020Serverless}. Generated functions are evaluated by running the project's existing test suites, complemented by code complexity metrics and CodeBLEU.

\noindent \textbf{Dynamic Service Generation.} Given a task description and optional runtime context, the model generates IoT service implementations. Evaluation measures semantic similarity between generated and reference code using CodeBERTScore \cite{Adnan2025ICSAC}.

\noindent \textbf{Traceability Link Recovery.} Given software architecture documentation (SAD) and a list of source code artifacts, the model must identify which code files implement each documented architectural element\cite{ClelandHuang2014Traceability, Soliman2021ICSA}. The dataset covers five open-source projects from the ArDoCo benchmark \cite{Fuchs2025ICSA}. Evaluation uses exact set matching with precision, recall, and F1.

\noindent \textbf{Microservice Generation.} Given requirements and codebase context for a microservice-based system, the model must generate complete, functional microservice implementations including controllers, services, data models, and configuration. The dataset covers 4 Java projects with 12 microservices total\cite{Adnan2026MicroservicesAgents}. Evaluation uses the project's unit and integration test suites alongside code complexity metrics.

\section{Usage and Walkthrough}
\label{sec:usage}
\begin{figure}[t]
    \centering
    \includegraphics[width=0.9\linewidth]{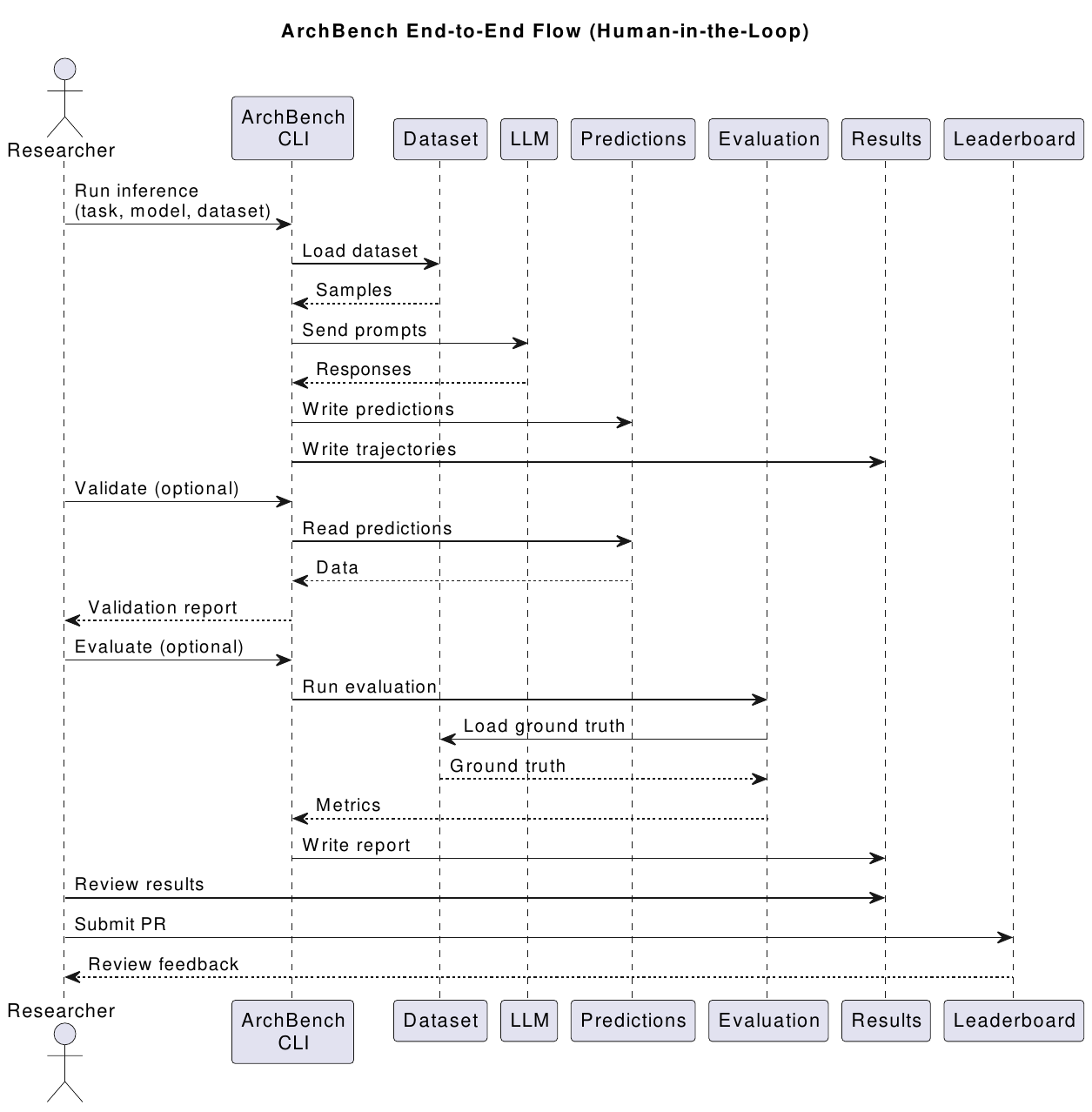}
    \caption{End-to-end ArchBench workflow. A researcher invokes the CLI to run inference, which loads the dataset, sends prompts to an LLM, and writes predictions with full trajectory logs. Evaluation against ground truth is run optionally within the same command. Results can then be submitted to the public leaderboard via pull request.}
    \label{fig:sequence}
\end{figure}
Fig.~\ref{fig:sequence} illustrates the end-to-end workflow. A researcher begins by installing the CLI tool via pip and setting an API key for their chosen LLM provider. Datasets are downloaded automatically on first use and cached locally, so no manual data preparation is required. A typical session involves a single command that runs inference and evaluation together:

\begin{lstlisting}[language=bash]
archbench inference --task adr --model gpt-4 \
    --output_dir results/ --evaluate
\end{lstlisting}

This command loads the task dataset, assembles prompts using the task's templates, sends them to the specified LLM, and writes structured predictions to a JSONL file. When the evaluate flag is set, the pipeline automatically runs evaluation against the ground truth and produces a report. Each run also generates a full trajectory log for reproducibility.

The output directory contains:
\begin{itemize}
  \item predictions.jsonl -- one prediction per line with instance ID, model output, and raw response
  \item report.json -- aggregated metrics with per-metric mean, standard deviation, min, and max
  \item trajectories/ -- individual JSON files recording each inference step
\end{itemize}

To contribute results to the public leaderboard, researchers submit a pull request to the website repository containing their evaluation report. This follows the submission model established by existing benchmarking leaderboards such as SWE-bench~\cite{Jimenez2024SWEbench}, where community contributions are vetted through open pull requests. Maintainers review the submission for completeness and validity and, upon acceptance, the results appear on the leaderboard alongside existing entries.

Adding a new task follows the same pull request workflow. The plugin architecture requires no changes to the core framework. A contributor creates a new task module containing a dataset loader (to fetch and parse data from an external source), prompt templates, a response parser, and evaluation logic with task-appropriate metrics. The module is registered in the CLI configuration, and a corresponding entry with task metadata is added to the web repository. This design keeps each task self-contained, so contributors only need to implement their task-specific logic while the framework handles inference orchestration, trajectory logging, and report generation.

\subsection*{Replication Package}
\label{sec:replication}

The ArchBench CLI and web platform are available as open-source repositories\footnote{CLI: https://github.com/sa4s-serc/archbench-cli}\footnote{Web: https://github.com/sa4s-serc/archbench}. The CLI can be installed via pip and supports optional dependency groups for evaluation metrics, BERTScore, and LLM inference providers. An archived release with a persistent DOI is available on Zenodo\footnote{Zenodo: https://doi.org/10.5281/zenodo.18635166}. All datasets used by the current tasks are publicly accessible and are fetched automatically by the tool. The replication package is licensed under Creative Commons Attribution 4.0 International (CC BY 4.0).

%
%
%
%

\section{Discussion}
\label{sec:discussion}

\noindent \textbf{Motivation and gap:} The software architecture community currently lacks a uniform set of datasets and a shared benchmarking platform for evaluating GenAI capabilities on architectural tasks. A recent multivocal literature review by Esposito et al.~\cite{ESPOSITO2026} found that rigorous testing of GenAI outputs was typically missing from existing studies, and identified the lack of architecture-specific datasets and the absence of sound evaluation frameworks as key open challenges. Individual studies rely on their own datasets, evaluation setups, and metrics, making cross-study comparison difficult~\cite{Dhar2024ICSA}. ArchBench is a direct effort to address these gaps by providing a centralized platform where datasets, evaluation pipelines, and results from disparate studies can be aggregated under a common infrastructure.

\noindent \textbf{Benefits and vision:} The broader AI and software engineering communities have advanced rapidly in part due to the availability of shared datasets and common evaluation platforms. The software architecture community currently lacks this shared infrastructure. ArchBench aims to fill this role by serving multiple stakeholders. For researchers, it provides a common ground to accelerate work on GenAI for architectural tasks, enabling direct comparison of approaches and reproducible evaluation. For practitioners, it offers a way to understand which LLMs or agentic approaches perform best on specific architectural tasks, informing tool adoption decisions. By creating a platform where researchers contribute tasks and evaluations while practitioners consume results, ArchBench bridges the gap between research and practice. Looking ahead, the availability of standardized architectural datasets and evaluation pipelines can also support the development and fine-tuning of models specifically tailored to software architecture tasks.

\noindent \textbf{Current status:} ArchBench currently hosts five architectural tasks with results from multiple LLMs on each. Two tasks (ADR Generation and Traceability Link Recovery) have fully automated evaluation pipelines in the CLI, meaning a researcher can go from dataset download to a scored report in a single command. The remaining three tasks have verified results from their original studies displayed on the leaderboard; their evaluation pipelines are being integrated into the CLI as the required test environments and metrics are standardized.

\noindent \textbf{Research questions enabled:} ArchBench enables research questions that were previously difficult to investigate, such as how architectural reasoning varies across model families, whether performance on one task correlates with another, and how prompting strategies affect output quality. The trajectory logging also enables fine-grained analysis of where models fail in multi-step architectural reasoning.

\noindent \textbf{Limitations:} ArchBench inherits the evaluation methodologies of its source datasets. Some tasks rely on automated metrics (ROUGE, BERTScore) that may not fully capture architectural quality, while others use test-suite pass rates that measure functional correctness but not design quality. The platform currently supports direct API-based evaluation and does not yet provide standardized sandboxed environments for agent-based tasks that require tool use. Additionally, the current task set, while diverse, does not cover all areas of software architecture; areas such as architectural smell detection, dependency analysis, and large-scale refactoring planning remain open for future contributions.

\noindent \textbf{Future directions:} We plan to expand ArchBench by integrating additional tasks from the community, adding support for agent-based evaluation environments, and incorporating qualitative evaluation via LLM-as-judge. We also plan to establish a regular evaluation cycle where new models are benchmarked as they are released, keeping the leaderboard current. As GenAI solutions become increasingly autonomous in proposing architectural changes, having a standardized platform to evaluate their reasoning becomes essential. Contributions from the community, both new tasks and new model evaluations, are welcomed through the open-source repositories.

%

\section*{Acknowledgment}
We thank the researchers who contributed benchmark tasks, shared datasets, and submitted evaluation results to the ArchBench platform. We also thank the contributors who helped improve the tool through code contributions and feedback. The platform is publicly available at \url{https://www.sabench.com/}.


\end{document}